\documentclass[12pt,prd,nofootinbib]{revtex4}
\usepackage{graphicx}
\usepackage{epsfig}
\usepackage{amsfonts}
\usepackage{amssymb}
\usepackage{amsmath}
\usepackage{bm}
\usepackage{latexsym}
\newcommand{\beq}{\begin{eqnarray}}
\newcommand{\eeq}{\end{eqnarray}}
\newcommand{\be}{\begin{equation}}
\newcommand{\ee}{\end{equation}}
\def\la{\mathrel{\mathpalette\fun <}}
\def\ga{\mathrel{\mathpalette\fun >}}
\def\fun#1#2{\lower3.6pt\vbox{\baselineskip0pt\lineskip.9pt
\ialign{$\mathsurround=0pt#1\hfil ##\hfil$\crcr#2\crcr\sim\crcr}}}

\newcommand{{\SD}}{\rm SD}

\newcommand{\vep}{\bm p}

\newcommand{\lsim}{\lower.7ex\hbox{$\;\stackrel{\textstyle<}{\sim}\;$}}
\begin{document}

\title{The vector coupling $\alpha_{\rm V}(r)$ and the scales $r_0,~r_1$
in the background perturbation theory}
\author{\firstname{A.M.}~\surname{Badalian}}
\email{badalian@itep.ru}
\affiliation{State Research Center, Institute of Theoretical and
Experimental Physics, Moscow 117218, Russia}
\author{\firstname{B.L.G.}~\surname{Bakker}}
\email{b.l.g.bakker@vu.nl} \affiliation{Department of Physics and Astronomy,
Vrije Universiteit, Amsterdam, The Netherlands}
\date{\today}
\begin{abstract}
We study the universal static potential $V_{\rm st}(r)$ and the
force, which are fully determined by two fundamental parameters:
the string tension $\sigma=0.18\pm 0.02$~GeV$^2$ and the QCD
constants $\Lambda_{\overline{\rm MS}}(n_f)$, taken from pQCD,
while the infrared (IR) regulator $M_{\rm B}$ is taken from the
background perturbation theory and expressed via the string
tension. The vector couplings $\alpha_{\rm V}(r)$ in the static
potential and $\alpha_{\rm F}(r)$ in the static force, as well as
the characteristic scales, $r_1(n_f=3)$ and $r_0(n_f=3)$, are
calculated and compared to lattice data. The result
$r_0\Lambda_{\overline{\rm MS}}(n_f=3)=0.77\pm 0.03$, which agrees
with the lattice data, is obtained for $M_{\rm B}=(1.15\pm
0.02)$~GeV. However, better agreement with the bottomonium
spectrum is reached for a smaller $\Lambda_{\overline{\rm
MS}}(n_f=3)=(325\pm 15)$~MeV and the frozen value of
$\alpha_V=0.57\pm 0.02$. The mass splittings $\bar M(1D)-\bar
M(1P)$ and $\bar M(2P)-\bar M(1P)$ are shown to be sensitive to
the IR regulator used. The masses $M(1\,^3D_3)=10169(2)$~MeV and
$M(1\,^3D_1)=10155(3)$~MeV are predicted.
\end{abstract}

\maketitle

\section{Introduction}
\label{sec.I}

The Hamiltonian formalism may be considered as a powerful tool to
study such hadron properties as meson spectroscopy, including high
excitations, hyperfine and fine-structure splittings of different
meson multiplets, leptonic widths, and radiative and strong meson
decays. For decades, different phenomenological Hamiltonians were
used in constituent quark models, and some of them were rather
successful in predictions of meson properties for low-lying states
\cite{ref.1,ref.2,ref.3,ref.4,ref.5}. However, in such models the
quark-antiquark potentials contain a large number of arbitrary
parameters like constituent quark masses, variable values of the
string tension and the QCD constant $\Lambda$, as well as an
overall additive fitting constant. Meanwhile, the relativistic
string Hamiltonian (RSH) $H_{\rm R}$, which was derived from the
gauge-invariant meson Green's function with the use of the QCD
Lagrangian \cite{ref.6}, contains a minimal set of fundamental
parameters: the current-quark masses, the string tension $\sigma$
fixed by the slope of the Regge trajectories for light mesons, and
the QCD constant $\Lambda(n_f)$, which can be taken from
perturbative QCD (pQCD). It is important that in the RSH the
spin-independent static potential $V_{\rm st}(r)$ is universal
and applicable for different $q_1\bar q_2$ mesons with arbitrary
masses (including $m_q=0$). This potential is defined via the
vacuum average over the Wilson loop $\langle W(C)\rangle$
\cite{ref.6,ref.7,ref.8,ref.9,ref.10,ref.11} and the 
only approximation made is that
$\langle W(C)\rangle$ is taken in the form of the minimal area
law, which appears to be a good approximation for separations
$r\ga T_g\sim 0.15$~fm, where $T_g$ is the vacuum correlation
length \cite{ref.8}.

The nonperturbative (NP) part of the static potential was shown to
have a linear behavior beginning at the separations $r\ga 0.2$~fm,
while at short distances, $r\la 0.15$~fm, the NP potential appears
to be proportional to $r^2$ \cite{ref.10}. Such a deviation from
linear behavior, in a very narrow region, gives a small effect for
all mesons, with the exception of $\Upsilon(1S)$, which has a
small size, $R\sim 0.20$~fm, and for which such a correction to
the confining potential should be taken into account.

In contrast to the NP part, the gluon-exchange (GE) part of the
static interaction is poorly defined on a fundamental level, with
the exception of the perturbative region valid at small distances.
In some models the GE potential depends on the renormalization scheme
(RS) and in the strong coupling $\alpha_{\rm V}(\mu)$ the scale
$\mu$, as well as the QCD constant $\Lambda$ chosen, may be
different for different quarks (mesons) \cite{ref.12}. Such
potentials violate the property of universality.

Moreover, there now is no consensus about the true value of the vector
coupling in the infrared (IR) region, which in phenomenological models
may vary in wide range \cite{ref.1,ref.2,ref.3,ref.4,ref.5}. 
Still the universality
of the GE potential was demonstrated in Ref.~\cite{ref.3}, where the
gross features of all mesons, from light to heavy, were successfully
described taking a phenomenological GE term with the frozen vector
coupling, called $\alpha_{\rm crit}$, equal to 0.60. A similar value of
$\alpha_{\rm crit}\sim 0.60$ was obtained in Ref.~\cite{ref.13} for a
more realistic GE interaction.

Unfortunately, existing lattice data on the static potential, which
is defined via the field-strength correlators, do not help to fix
$\alpha_{\rm crit}$. Moreover, if at $r\ga 0.20$~fm the lattice
static potential is parametrized as in the string theory, $V_{\rm
lat}(r)=\sigma r -\frac{e}{r}$, then in $SU(3)$ lattice QCD the Coulomb
constant $e$ appears to be small: $e=0.40$, or $\alpha_{\rm lat}=0.30$,
in Ref.~\cite{ref.14}, and even a smaller number, $e\simeq 0.30$, or
$\alpha_{\rm lat}\simeq 0.22$, was calculated in Ref.~\cite{ref.15}.

A second difference with the lattice results is about the
$r$-dependence of the strong coupling: on the lattice the
saturation of the vector coupling occurs already at small
distances, $r\sim 0.20$~fm, while it takes place at significantly
larger separations, $r\ga 0.60$~fm, if one uses the vector
coupling derived within the background perturbation theory (BPT)
\cite{ref.13, ref.16}. Therefore it is of a special importance to
compare lattice results and ours for the first and second
derivatives of the static potential.
For the static force we calculate here its characteristic scales
$r_0$ and $r_1$, while the second derivative of the potential,
which does not depend on the NP part, allows to determine the
derivatives of the vector coupling.

Our approach has the following features:
\begin{enumerate}
\item The asymptotic freedom (AF) behavior of the vector
coupling $\alpha_{\rm V}(q)$ at large momenta $q^2$ is taken into
account.  This coupling is defined by the ``vector" QCD constant
$\Lambda_{\rm V}(n_f)$, which is directly expressed through the
conventional $\Lambda_{\overline{\rm MS}}(n_f)$ due to results of
Refs.~\cite{ref.17}. The values of $\Lambda_{\overline{\rm MS}}(n_f)$
are considered to be known from pQCD \cite{ref.18, ref.19}.
\item We do not use here the correspondence $q=1/r$, since it is
valid only at small $r$ \cite{ref.20, ref.21}.
\item The IR regulator $M_{\rm B}$ is taken from Ref.~\cite{ref.16},
where $M_{\rm B}$ is shown to be determined by the string tension according
to the relation: $M_{\rm B}^2=2\pi\sigma$.
\item In the vector coupling two-loop corrections are taken into
account, giving a contribution $\sim 30\%$, while the higher
terms, dependent on the RS, are omitted, in agreement with the concept
of Shirkov \cite{ref.22}.
\end{enumerate}
In the framework of our approach the values of the frozen coupling
may be fixed with $\sim 10\%$ accuracy.

\section{The static potential}
\label{sec.II}

The universal quark-antiquark potential contains the linear
confining term and the GE part:
\begin{equation}
 V_{\rm st}(r)= \sigma r + V_{\rm GE}(r),
\label{eq.1}
\end{equation}
and has the property of additivity at $r\leq 1.0$~fm, which is confirmed
by the Casimir scaling, studied analytically \cite{ref.11} and numerically
on the lattice \cite{ref.23}.  Here the string tension is not an arbitrary
parameter, but fixed by the slope of the Regge trajectories of the light
mesons, which is known to good accuracy, $\sigma=0.180\pm 0.002$~GeV$^2$.

In the GE potential Eq.~(\ref{eq.1})
\begin{equation}
 V_{\rm GE}(r) = - \frac{4}{3}\frac{\alpha_{\rm V}(r)}{r},
\label{eq.2}
\end{equation}
the vector coupling in coordinate space $\alpha_{\rm V}(r)$,
is defined through the vector coupling $\alpha_{\rm V}(q^2)$ in
momentum space as follows,
\begin{equation}
 \alpha_{\rm V}(r)=\frac{2}{\pi}\int\limits_0^\infty {\rm d}q
 \frac{\sin(qr)}{q}\,\alpha_{\rm V}(q^2).
\label{eq.3}
\end{equation}
For large $q^2$ there exists an important relation between
$\alpha_{\rm V}(q^2)$ in momentum space and the conventional
$\alpha_{\rm s}(q^2)$ in the $\overline{\rm MS}$ RS \cite{ref.17}.
In pQCD the cross sections and other observables are predicted in
terms of this coupling. The coupling $\alpha_{\rm s}(q^2)$ is
measured at different (large) energy scales $q^2$ and the values
obtained are usually presented at a common energy scale, equal to
the $Z$-boson mass, $M_Z=91.188$~GeV.
From numerous experimental studies, like the hadronic widths of
the $Z^0$ boson, the $\tau$-lepton decays, radiative
$\Upsilon(1S)$ decays, jet production in $e^+e^-$ annihilation,
and the structure functions in deep inelastic scattering, the
world average value of the strong coupling is now determined with
a good accuracy, $\alpha_{\rm s}(m_Z)=0.1184\pm 0.0007$
\cite{ref.18, ref.19}. As a consequence, the QCD constant
$\Lambda_{\overline{\rm MS}}(n_f=5)$ is now known with good
accuracy. Then, using the matching procedure at the quark mass
thresholds, the other $\Lambda_{\overline{\rm MS}}(n_f)$ for
$n_f=3,4$ are calculated and the three-loop calculations give the
following $\Lambda_{\overline{\rm MS}}(n_f)$ \cite{ref.19}:
\begin{eqnarray}
 \Lambda_{\overline{\rm MS}}(n_f=3) & = & (339\pm 10)~{\rm MeV},
\nonumber \\
 \Lambda_{\overline{\rm MS}}(n_f=4) & = & (296\pm 10)~{\rm MeV},
\nonumber \\
 \Lambda_{\overline{\rm MS}}(n_f=5) & = & (213\pm 8)~{\rm MeV}.
\label{eq.4}
\end{eqnarray}
These numbers can be used to define the ``vector" constants
$\Lambda_{\rm V}(n_f)$, expressed via $\Lambda_{\overline{\rm MS}}(n_f)$
\cite{ref.17} (see also below Sect.~\ref{sec.IV}). They appear to be
significantly larger, e.g. $\Lambda_{\rm V}(n_f=3)=(500\pm 15)$~MeV
corresponds to the value $\Lambda_{\overline{\rm MS}}(n_f=3)=(339\pm
10)$~MeV from Eq.~(\ref{eq.4}).

The analysis of $V_{\rm GE}(r)$ shows that perturbative effects
determine $\alpha_{\rm V}(r)$ only at very small distances $r\la 0.06$~fm
\cite{ref.20} and this result was confirmed by the lattice measurements
of the static potential \cite{ref.21}.

In BPT this potential is defined in the presence of the background
fields and therefore cannot be considered like the one-gluon-exchange
interaction. Moreover, in this GE term the NP effects become
important, beginning from very short distances, and our goal here
is to determine the vector coupling in the IR region.

For heavy quarkonia, the importance of NP effects was understood
already in 1975, just after the discovery of the charmed quark, when
the Cornell group introduced the linear + Coulomb potential with a rather
large vector coupling, $\alpha_{\rm V}=\rm constant=0.39$ \cite{ref.1}
over the whole region, neglecting the AF behavior. However, future studies
have shown that the AF behavior of the vector coupling is very important,
in particular, for the wave functions (w.f.) and its derivatives at the
origin \cite{ref.24, ref.25}. Later it has become clear that if the
AF effect is taken into account, then the frozen value of $\alpha_{\rm V}$
becomes larger \cite{ref.3, ref.13}.

On the fundamental level, not many theoretical attempts were undertaken
to determine the strong coupling in the IR region, although on the
phenomenological level a regularization of the strong coupling was
suggested long ago, with the prescription to introduce the IR regulator
into the logarithm $\ln \frac{q^2}{\Lambda^2}$, changing it into $\ln
\left(\frac{q^2+M_{\rm 2g}^2}{\Lambda^2}\right)$ \cite{ref.26}.  This IR
regulator was interpreted as an effective two-gluon mass $M_{\rm 2g}=2m_g$
with the mass $m_g\sim 0.50$~GeV. However, in QCD the appearance of
the gluon mass is forbidden by gauge invariance and the meaning and the
value of the IR regulator remained unsolved for many years.

Recently, within BPT just the same type of logarithm, as in
Ref.~\cite{ref.26}, was derived and the IR regulator (denoted as
$M_{\rm B}$) was shown to be expressed through the string tension
\cite{ref.16}. Thus the IR regulator $M_{\rm B}$ is not an
additional parameter and has the meaning of the mass of the
two-gluon system, connected by the fundamental string (white
object). Its value is determined by the equation: $M_{\rm
B}^2=2\pi\sigma$, giving $M_{\rm B}\simeq (1.06\pm 0.11)$~GeV for
$\sigma\sim 0.180$~GeV$^2$, where the accuracy of the calculations
is determined by the accuracy of the WKB method used ($\sim
10\%$).

The IR regulator was also studied in so-called ``massive" pQCD,
developed within Analytic Perturbation Theory, and the predicted
value is obtained in the range $(0.9-1.2)$~GeV \cite{ref.22}.
However, admissible variations of the regulator $M_{\rm B}$ in the
range $1.0-1.15$~GeV give rise to significant differences in the
frozen value $\alpha_{\rm crit}$, which is the same in the
momentum and the coordinate spaces: $\alpha_{\rm
V}(q=0)=\alpha_{\rm V}(r\to \infty)= \alpha_{\rm crit}(n_f=3)$.
For example, taking the central value of $\Lambda_{\overline{\rm
MS}}(n_f=3)=339$~MeV from Eq.~(\ref{eq.4}) and the corresponding
$\Lambda_{\rm V}(n_f=3)= 1.4753\,\Lambda_{\overline{\rm
MS}}(n_f=3)=500$~MeV, one obtains $\alpha_{\rm crit}(2-{\rm
loop})$ equal to the large value 0.82 for $M_{\rm B}=1.0$~GeV and
a smaller value 0.635 for the larger $M_{\rm B}=1.15$~GeV.  Such
different critical values give different results for the meson
spectra and one needs to determine the IR regulator, as well as
$\Lambda_{\overline{\rm MS}}(n_f=3)$, with great accuracy. Notice
that the smaller value $\Lambda_{\overline{\rm MS}}(n_f=3)=(292\pm
29)$~MeV, as compared to the one in Eq.~(\ref{eq.4}), was used in
pQCD in Ref.~\cite{ref.27} and an even smaller value was used in
Ref.~\cite{ref.28}.

Here, as a test, we calculate the bottomonium spectrum and study
how it depends on the IR regulator and the value of $\Lambda_{\rm
V}(n_f=3)$ used. The frozen value is shown to be determined by the
ratio $\eta^2=\frac{M_{\rm B}^2}{\Lambda_{\rm V}^2}$ (or ${\tilde
\eta}^2= \frac{M_{\rm B}^2}{\Lambda_{\overline{\rm MS}}^2}$) and taking
$\Lambda_{\rm V}(n_f=3)\sim 500\pm 15$~MeV from Eq.~(\ref{eq.16}),
corresponding to the pQCD value given in Eq.~(\ref{eq.4}), we obtain
that $M_{\rm B}=(1.15\pm 0.02)$~GeV provides the best description of
the bottomonium spectrum, and this value agrees with the prediction
from Ref.~\cite{ref.16}. The value of the IR regulator may be smaller,
by $\sim 10\%$, if a smaller QCD constant is taken.

\section{Relativistic string Hamiltonian}
\label{sec.III}

We use here the the RSH  $H_{\rm R}$, which was derived from the
gauge-invariant meson Green's function, performing several steps
(see Refs.~\cite{ref.6,ref.9}). For a meson $q_1\bar q_2$ with the
masses $m_1$ and $m_2$ the RSH contains several terms,
\begin{equation}
 H_{\rm R}= H_0 + H_{\rm SD} + H_{\rm str} + H_{\rm SE},
\label{eq.5}
\end{equation}
where the part $H_{\rm SD}$ refers to the spin-dependent potential,
like hyperfine or fine-structure interactions; the term $H_{\rm str}$
comes from the rotation of the string itself and determines the so-called
string corrections for the states with $l\neq 0$, while $H_{\rm SE}$ comes
from the NP self-energy contribution to the masses of the quark and the
antiquark \cite{ref.29}. All these terms appear to be much smaller than
the unperturbed part $H_0$ (the same for all mesons), and therefore can
be considered as a perturbation. The part $H_0$ is derived in the form,
\begin{equation}
 H_0=\frac{\omega_1}{2} +\frac{\omega_2}{2} +\frac{m^2_1}{2\omega_1}+
 \frac{m^2_2}{2\omega_2} +\frac{\vep^2}{2\omega_{\rm red}}
 +V_{\rm st}(r).
\label{eq.6}
\end{equation}
Here the variables  $\omega_i$ are the kinetic energy operators,
which have to be determined from the extremum condition,
$\frac{\partial H_0}{\partial\omega_i} =0$, 
$1/\omega_{\rm red} = 1/\omega_1 + 1/\omega_2$, giving
\begin{equation}
 \omega_i = \sqrt{ m^2_i+ \vep^2}~~(i=1,2),
\label{eq.7}
\end{equation}
and therefore Eq.~(\ref{eq.6}) can  be rewritten as
\begin{equation}
 H_0 = \sqrt{m_1^2 + \vep^2} + \sqrt{m_2^2 + \vep^2} + V_{\rm st}(r).
\label{eq.8}
\end{equation}
The general form of the Hamiltonian $H_{\rm R}$ describes
heavy-light and light mesons, but it is very simplified for
bottomonium, which has the largest number of levels below the open
flavor threshold. Altogether there are nine $b\bar b$ multiplets
with $l=0,1,2,3$ and seven of them were already observed; just
this extensive information may be used to test a universal static
potential. An additional piece of information on the coupling
$\alpha_{\rm s}(\mu)$ at different scales $\mu$ may be extracted
from studies of the hyperfine and fine-structure effects in
bottomonium \cite{ref.30}.
By derivation, in the RSH the quark (antiquark) mass $m_i$ is equal to
the current quark (antiquark) mass, $\bar m_i(\bar m_i)$ in the
$\overline{\rm MS}$ RS, and therefore it is not a fitting
parameter. In the case of a heavy quark one needs to take into
account corrections perturbative in $\alpha_{\rm s}$, i.e., to use
the pole mass of a heavy quark, which is taken here to two-loop
accuracy:
\begin{equation}
 m_Q=\bar{m}_Q(\bar{m}_Q)
 \left\{ 1+\frac{4}{3}
 \frac{\alpha_{\rm s}(\bar{m}_Q)}{\pi}
 +\xi_2\Biggl (\frac{\alpha_{\rm s}}{\pi} \Biggr )^2
  \right \},
\label{eq.9}
\end{equation}
where $\xi_2$ may be taken from Ref.~\cite{ref.18}. For the $b$
quark the pole mass can symbolically be written as $m_b(\rm
pole)=\bar m_b(\bar m_b)(1+ 0.09 + 0.05)$, where the second and
third terms come from the $\alpha_{\rm s}$ and $\alpha_{\rm s}^2$
corrections. In our calculations $m_b({\rm pole})=(4.81\pm
0.03)$~GeV is used, which corresponds to the conventional current
mass $\bar m_b(\bar m_b)=(4.22\pm 0.03)$~GeV.

It is important that in bottomonium the calculated string and
self-energy terms are very small, $\leq 1$~MeV, and therefore the RSH
reduces to $H_{\rm R}=H_0+H_{\rm SD}$, as it follows from
Eq.~(\ref{eq.8}),

\begin{equation}
 H_0 = 2\sqrt{m_b^2+ \vep^2} + V_{\rm st}(r),
\label{eq.10}
\end{equation}
with a kinetic term similar to that in the spinless Salpeter
equation (SSE), which is often used in relativistic models with
constituent quark masses. Such a coincidence between the kinetic
terms in the SSE and the RSH, which was derived from first
principles, possibly explains the success of relativistic models with
this type of the kinetic term \cite{ref.2,ref.3}.

An important feature of the RSH is that it does not contain an
overall additive (fitting) constant, which is usually present in
models with constituent quark masses and also in the lattice
static potential \cite{ref.15}.  Notice that the presence of such
a constant in the meson mass violates the linear behavior of the
Regge trajectories for light mesons. On the contrary, with the use
of $H_{\rm R}$ linear Regge trajectories can be easily derived
with the correct slope and intercept \cite{ref.9}
($\sigma=0.180\pm 0.002$~GeV$^2$ was extracted from the slope of
the Regge trajectories for light mesons). In heavy quarkonia
low-lying states do not lie on linear Regge trajectories, because
of strong GE contributions.
The static potential present in $H_0$ is supposed to be a
universal one.

\section{The vector coupling in momentum space}
\label{sec.IV}

The vector coupling $\alpha_{\rm V}(q)$ in momentum space is taken
here in two-loop approximation, where the coupling does not depend
on the RS. Later, for $\alpha_{\rm V}(q^2)$ we shall use the
notation $\alpha_{\rm B}(q^2)$, bearing in mind that it contains
the IR regulator $M_{\rm B}$, determined as in BPT \cite{ref.16}:
\begin{equation}
 \alpha_{\rm B}(q^2)=\frac{4\pi}{\beta_0t_{\rm B}}
 \left(1-\frac{\beta_1}{\beta_0^2}\frac{\ln t_{\rm B}}{t_{\rm B}}\right).
\label{eq.11}
\end{equation}
Here $M_{\rm B}$, entering the logarithm $t_{\rm
B}=\ln\frac{\left(q^2+M_{\rm B}^2)\right)}{\Lambda_{\rm V}^2}$, is
not a new parameter but determined via the string tension
\cite{ref.16} in the fundamental representation:
\begin{equation}
 M_{\rm B}^2=2\pi\sigma
\label{eq.12}
\end{equation}
with $\sigma=(0.180\pm 0.002)$~GeV$^2$. The accuracy of the
relation (\ref{eq.12}) is determined by the accuracy of the WKB
approximation used in Ref.~\cite{ref.16}, which is estimated to be
$\leq 10\%$. Therefore
\begin{equation}
 M_{\rm B}=(1.06\pm 0.11)~{\rm GeV}.
\label{eq.13}
\end{equation}
The analysis of the bottomonium spectrum shows that the larger
values, $M_{\rm B}=(1.15-1.20)$~GeV, are preferable, if a large
$\Lambda_{\rm V}(n_f=3)=(500\pm 15)$~MeV, corresponding to the
pQCD value $\Lambda_{\overline{\rm MS}}(n_f=3)=(339\pm 10)$ from
Eq.~(\ref{eq.4}), is taken, while for the smaller $M_{\rm
B}=(1.05\pm 0.05)$~GeV and the same $\Lambda_{\rm V}$ one obtains
too large a $2P-1P$ splitting and also a large $b$-quark pole mass,
$m_b=4.90$~GeV.

The ``vector" constant $\Lambda_{\rm V}(n_f)$ may be expressed
through the conventional $\Lambda_{\overline{\rm MS}}(n_f)$, if
one uses the connection between the strong couplings in momentum
space and the $\overline{\rm MS}$ RS, established in
Ref.~\cite{ref.17}, which is valid at large $q^2$:
\begin{equation}
 \alpha_{\rm V}(q^2)=\alpha_{\rm s}(q^2)
 \left(1 + \frac{a_1}{4\pi}\alpha_{\rm s}(q)\right) \approx
 \frac{\alpha_{\rm s}(q)}{\left(1-\frac{a_1}{4\pi}\alpha_{\rm s}(q)\right)}.
\label{eq.14}
\end{equation}
Here $a_1=\frac{31}{3}- \frac{10}{9}n_f$. Notice, that the first
order correction in Eq.~(\ref{eq.14}) is important, otherwise
$\Lambda_V$ and $\Lambda_{\overline{\rm MS}}$ would be equal. For
our purpose it is enough to use in Eq.~(\ref{eq.14}) the one-loop
approximation for both couplings: $\alpha_{\rm
B}=\frac{4\pi}{\beta_0}\ln\frac{q^2}{\Lambda_{\rm V}^2}$ and
$\alpha_s=\frac{4\pi}{\beta_0}\ln\frac{q^2}{\Lambda_{\overline{\rm
MS}}^2}$ with $\beta_0=11 -\frac{2}{3}n_f$. Then from
Eq.~(\ref{eq.14}) the relation, $\ln\frac{q^2}{\Lambda_{\rm V}^2}=
\left(\ln\frac{q^2}{\Lambda_{\overline{\rm MS}}^2}
-\frac{a_1}{\beta_0}\right)$, follows and its solution is
\begin{equation}
 \Lambda_{\rm V}(n_f)=\Lambda_{\overline{\rm MS}}(n_f)
 \exp\left(-\frac{a_1}{2\beta_0}\right).
\label{eq.15}
\end{equation}
This relation gives
\begin{eqnarray}
 \Lambda_{\rm V}(n_f=3) & = & 1.4753~\Lambda_{\overline{\rm MS}}(n_f=3),
\nonumber \\
 \Lambda_{\rm V}(n_f=4) & = & 1.4238~\Lambda_{\overline{\rm MS}}(n_f=4),
\nonumber \\
 \Lambda_{\rm V}(n_f=5) & = & 1.3656~\Lambda_{\overline {MS}}(n_f=5).
\label{eq.16}
\end{eqnarray}
If one takes the perturbative $\Lambda_{\overline{MS}}(n_f)$ from
Eq.~(\ref{eq.4}), then the following values for the ``vector"
constants in pQCD are obtained:
\begin{eqnarray}
 \Lambda_{\rm V}(n_f=5) & = & (291\pm 11)~{\rm MeV},
\nonumber \\
 \Lambda_{\rm V}(n_f=4) & = & (421\pm 15)~{\rm MeV},
\nonumber \\
 \Lambda_{\rm V}(n_f=3) & = & (500\pm 15)~{\rm MeV}.
\label{eq.17}
\end{eqnarray}
Here $\Lambda_{\rm V}(n_f=5)$ as well as $\Lambda_{\overline{\rm
MS}}(n_f=5)$ are considered to be known with a good accuracy. 
The matching procedure is performed here for the coupling $\alpha_{\rm
B}(q^2)$ in momentum space, not for $\alpha_{\rm s}(q^2)$. It is
interesting to underline that in this case the calculated values
of $\Lambda_{\rm V}(n_f)$ for $n_f=4,5$ practically coincide with
their values in Eq.~(\ref{eq.17}), although now the IR regulator
is taken into account.

We use here two sets of $\Lambda_{\rm V}(n_f)$ for two different
values of $M_{\rm B}$, equal to 1.15 GeV and 1.0 GeV, respectively. 
Then for $M_{\rm B}=1.15$~GeV
\begin{equation}
 \Lambda_{\rm V}(n_f=5)=310~{\rm MeV},~~
 \Lambda_{\rm V}(n_f=4)=429.6~{\rm MeV},~~
 \Lambda_{\rm V}(n_f=3)=497.4~{\rm MeV},
\label{eq.18}
\end{equation}
and if $M_{\rm B}=1.00$~GeV,
\begin{equation}
 \Lambda_{\rm V}(n_f=5)=315~{\rm MeV}, ~~
 \Lambda_{\rm V} (n_f=4)=435~{\rm MeV}, ~~
 \Lambda_{\rm V}(n_f=3)=499.7~{\rm MeV}.
\label{eq.19}
\end{equation}
Thus the fitted values of $\Lambda_{\rm V}(n_f)$ weakly depend on
the IR regulator $M_{\rm B}$, if it is taken in the range
$1.0-1.15$~GeV, varying within $\pm 5$~MeV. (Here the matching
was performed at the quark mass thresholds: $q_{54}=4.20$~GeV and
$q_{43}=1.50$~GeV). The difference between these two sets becomes
manifest only in the frozen value: $\alpha_{\rm
crit}(q=0,n_f=3)=0.630$ for $M_{\rm B}=1.15$~GeV and $\alpha_{\rm
crit}=0.819$ for $M_{\rm B}=1.0$~GeV.

\begin{figure}[htb]
\vspace{3ex}
\begin{center}
 \includegraphics[width=60mm]{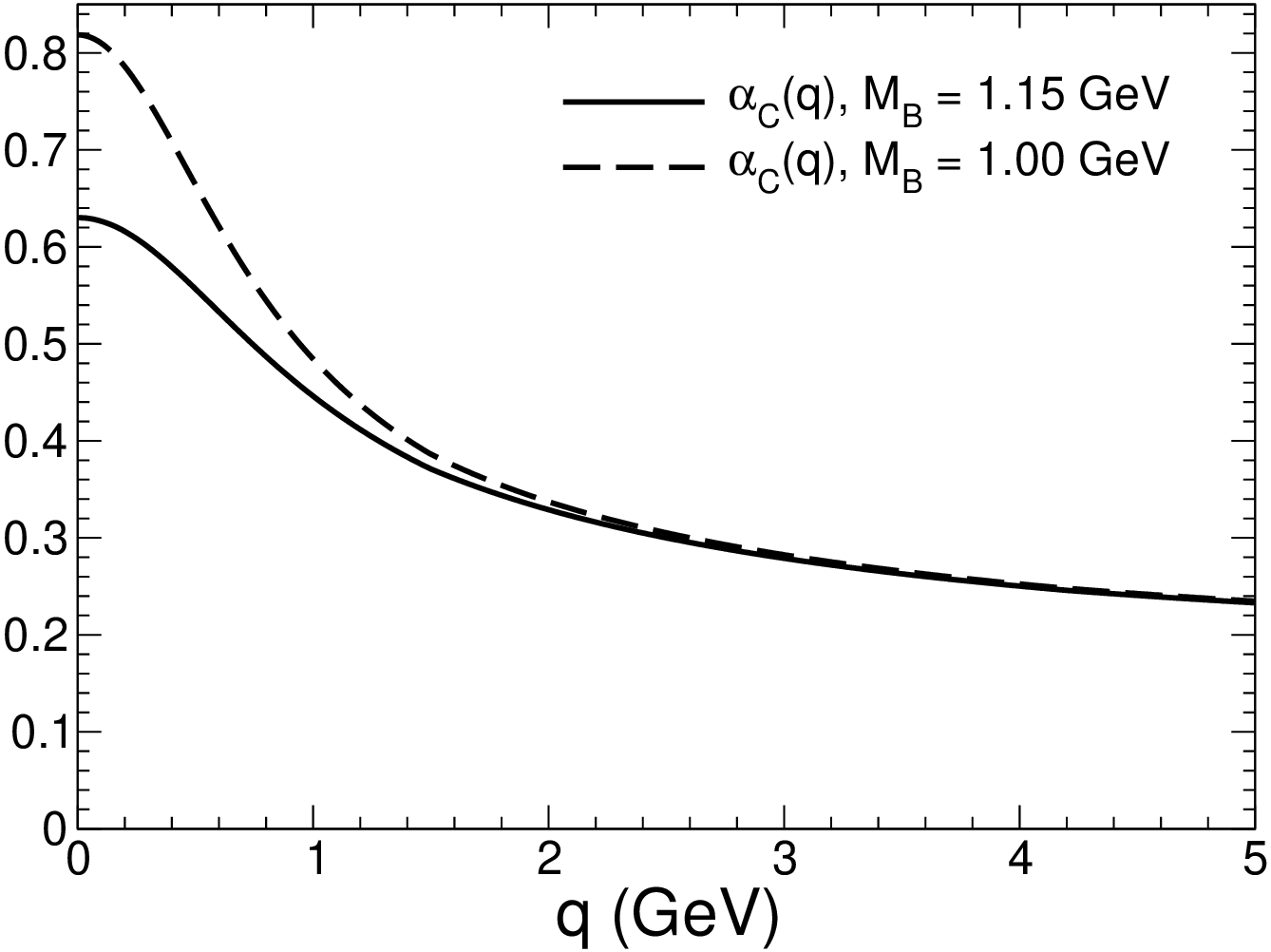}
\caption{Compound $\alpha_{\rm B}(q)$ for $M_{\rm B} = 1.15$ GeV and
$\Lambda_V(n_f)$ from Eq.~(\ref{eq.18}) and for $M_{\rm B} = 1.00$ GeV and
$\Lambda_V(n_f)$ from Eq.~(\ref{eq.19}).
\label{fig.1}}
\end{center}
\end{figure}

In Fig.~\ref{fig.1} we give two curves for the compound
$\alpha_{\rm B}(q^2)$, with almost the same parameters
$\Lambda_{\rm V}$ from Eqs.~(\ref{eq.18}) and (\ref{eq.19}), but
significantly different frozen couplings owing to the change of the IR
regulator by $15\%$.
Later, for a comparison we shall also use the set with a smaller
$\Lambda(n_f=3)=465$~MeV,
 \begin{equation}
  \Lambda_V(n_f=5)=292~ {\rm MeV},~~ \Lambda_V(n_f=4)=406~{\rm MeV},~~
  \Lambda(n_f=3)= 465~{\rm MeV},
 \label{eq.20}
 \end{equation}
and a smaller $\alpha_{\rm crit}(q^2=0, n_f=3)=0.5712$.

Notice that the value of $\alpha_{\rm crit}=0.57$ is close to those
which were used in phenomenology \cite{ref.2,ref.3,ref.4,ref.9},
with typical values $\alpha_{\rm crit}\sim 0.54-0.60$, but is
significantly larger than the lattice $\alpha({\rm lat})\sim
(0.22-0.30)$ in full QCD \cite{ref.14,ref.15}. The reason for that
discrepancy possibly comes from lattice artifacts, present in the
lattice GE potential \cite{ref.15}, and also from an additional
normalization condition, usually put on the lattice static potential
\cite{ref.27,ref.31}.

\begin{figure}
 \includegraphics[width=60mm]{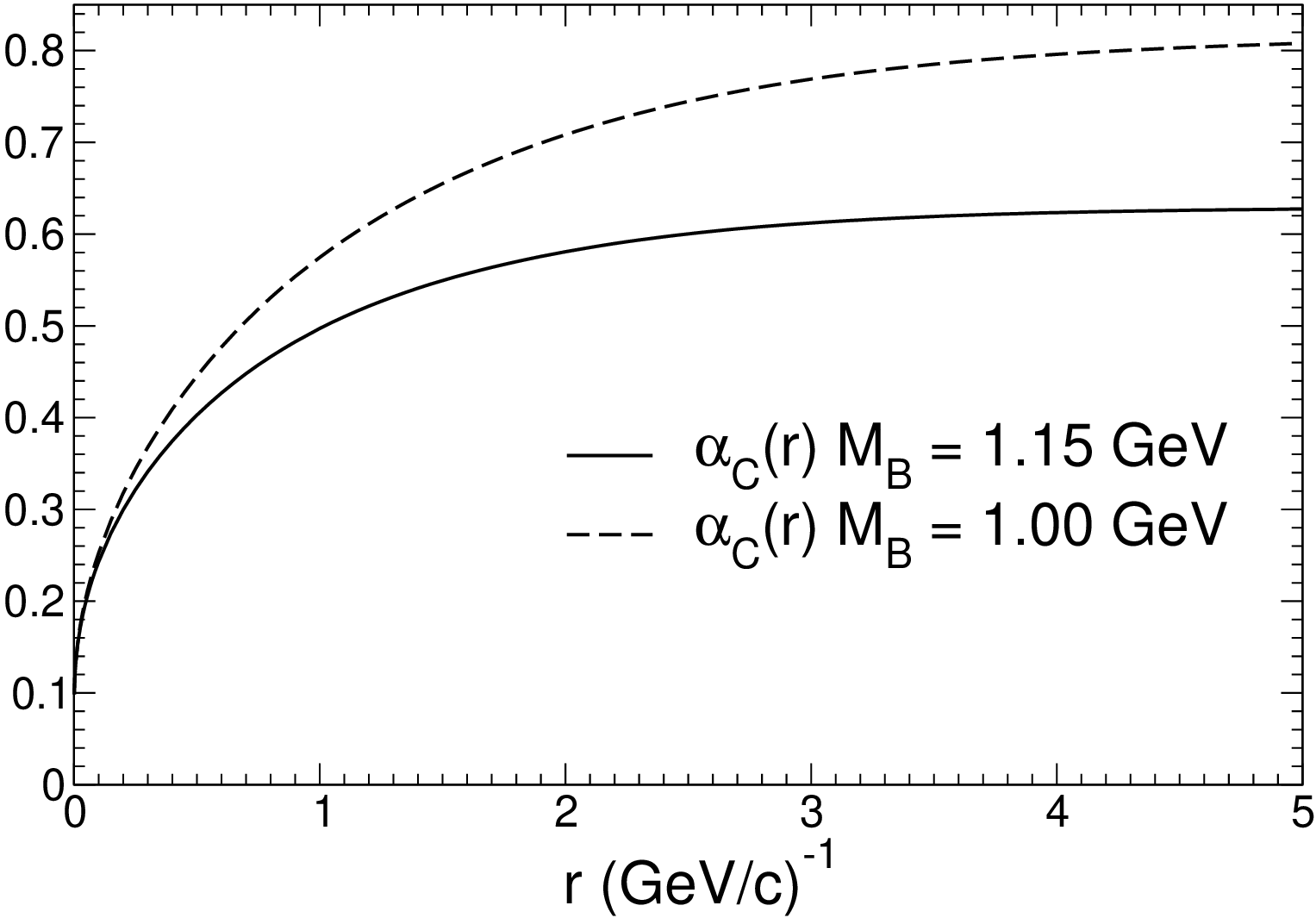}
\caption{Compound $\alpha_{\rm B}(r)$ with $\alpha_{\rm crit} =0.63015$
and the parameters from Eq.~(\ref{eq.18}) (solid line), and $\alpha_{\rm B}(r)$
(dashed line) with $\alpha_{\rm crit}=0.819$ and the parameters from
Eq.~(\ref{eq.19}).\label{fig.2}}
\end{figure}

\begin{figure}
\vspace{1ex}
 \includegraphics[width=60mm]{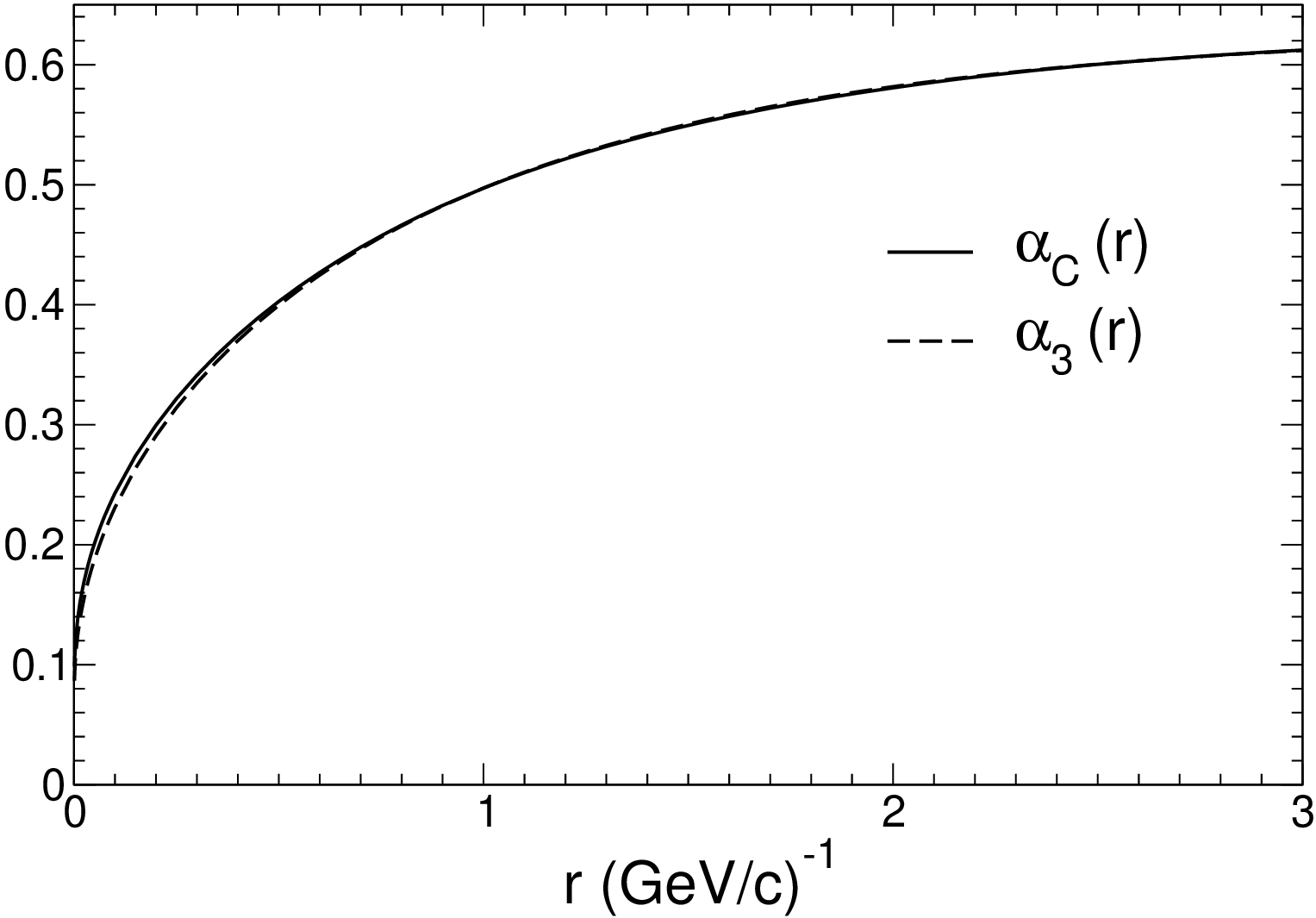}
\caption{Comparison of the compound $\alpha_{\rm B}(r)$ with
parameters from Eq.~(\ref{eq.18}) and $\alpha_{\rm B}(r)$
with fixed $n_f=3$ and the same $\alpha_{\rm crit}$ and
$\Lambda_V(n_f=3)=0.4974$~GeV. \label{fig.3}}
\end{figure}

Also, in contrast to some lattice potentials, where saturation of
the vector coupling takes place at very small distances, $r\sim
0.2$~fm \cite{ref.14,ref.15}, in our approach the vector coupling
is approaching its critical value at the much larger distances $r\ga 0.6$~fm
(see Figs.~\ref{fig.2},\ref{fig.3}).

From Eq.~(\ref{eq.17}) it is evident that the asymptotic coupling
$\alpha_{\rm crit}$ is fully determined by the ratio
$\eta^2=\frac{M_{\rm B}}{\Lambda_{\rm V}^2}$ and in two-loop
approximation is given by
\begin{equation}
 \alpha_{\rm B}(q=0) =\frac{4\pi}{\beta_0t_0}\left(1-\frac{\beta_1}{\beta_0^2}
 \frac{\ln t_0}{t_0}\right),
\label{eq.21}
\end{equation}
with the logarithm
\begin{equation}
 t_0=\ln \eta^2= \ln \left(\frac{M_{\rm B}^2}{\Lambda_V^2}\right).
\label{eq.22}
\end{equation}
It is clear that to determine the frozen coupling with great
accuracy, one needs to exclude small uncertainties in the values
of $\Lambda_V(n_f=3)$ and $M_{\rm B}$, which can change
$\alpha_{\rm crit}(n_f=3)$ by $\sim 30\%$.

It is also important that the critical couplings in the momentum
and the coordinate spaces coincide:
\begin{equation}
   \alpha_{\rm B}({\rm crit})=\alpha_{\rm B} (r\to \infty) =\alpha_{\rm
   B}(q=0).
\label{eq.23}
\end{equation}
It is of interest to understand why in phenomenological models the
smaller values $\Lambda_V(n_f=3)\sim 330-380$~MeV are often used
(compared to those from Eq.~(\ref{eq.17})), giving, nevertheless,
a good description of the low-lying meson states. Such values of
$\Lambda_V$ correspond to a smaller $\Lambda_{\overline{\rm
MS}}(n_f=3)\sim 250$~MeV, as compared to that from
Eq.~(\ref{eq.4}), and are close to those calculated in the
quenched approximation on the lattice, $\Lambda(n_f=0)=(245\pm
20)$~MeV \cite{ref.32}.  Nevertheless, in this case a reasonable
agreement with experiment is also reached due to a smaller value
taken for $M_{\rm B}$, so that the frozen  constant is again
large, $\alpha_{\rm B}(\rm crit)\sim 0.60$.
Later we will show that the lattice results appear to be in a
better agreement with ours, when the static force and the second
derivative of the static potential are compared.

\section{The vector coupling in coordinate space}
\label{sec.V}

The vector coupling in the coordinate space is defined according
to Eq.~(\ref{eq.3}), where the integral can be rewritten in a
different way, introducing the variable $y=\frac{q}{\Lambda_V}$,
\begin{equation}
 \alpha_{\rm B}(r \Lambda_V, \eta^2)=\frac{2}{\pi}\int\limits_0^\infty dy
 \frac{\sin(r \Lambda_V y)}{y}\alpha_{\rm B}(y, \eta^2),
\label{eq.24}
\end{equation}
This expression explicitly shows that  $\alpha_{\rm B}(r)$ depends
on the combination $r \Lambda_V(n_f)$, if $n_f$ is fixed, and also
on the parameter $\eta^2=\frac{M_{\rm B}^2}{\Lambda_V(n_f)^2}$.

In Fig.~2 two couplings $\alpha_{\rm B}(r)$ are shown for two sets
of $\Lambda(n_f)$ from Eqs.~(\ref{eq.18}), (\ref{eq.19}), where
the critical values are equal to 0.630 ($M_{\rm B}=1.15$~GeV) and and
0.819 ($M_{\rm B}=1.0$~GeV).

In Fig.~3 the calculated ``compound" $\alpha_{\rm B}(r)$ with
$M_{\rm B}=1.15$~GeV is compared to the coupling $\alpha_{\rm
B}(r)$, in which $n_f=3$ is fixed (no matching), while for both
couplings their critical values coincide and are equal to 0.630.
As seen from Fig.~3, both curves are very close to each other and,
perhaps just owing to this fact, the vector coupling with fixed
$n_f=3$ may be used in phenomenological models. It also indicates
that the frozen value of the coupling is of primary importance.

Notice that the situation is different for light and strange
mesons, which have large sizes, and for them a screening of the GE
interaction is possible, which can occur owing to open channels,
decreasing the vector coupling.

\section{The static force and the function $c(r)$}
\label{sec.VI}

To have an additional test of the calculated vector coupling $\alpha_{\rm B}(r)$
we consider here the static force,
\begin{equation}
 F_{\rm B}(r) =V_{\rm st}'(r) =\sigma + V'_{\rm GE}(r)
 \equiv \sigma + \frac43 \frac{\alpha_{\rm F}(r)}{r^2},
\label{eq.25}
\end{equation}
where the coupling
\begin{equation}
 \alpha_{\rm F}(r) = \alpha_{\rm B}(r) - r \alpha'_{\rm B}(r)
\label{eq.26}
\end{equation}
is introduced. The coupling $\alpha_{\rm F}(r)$ is smaller than
$\alpha_{\rm B}(r)$, since the derivative $\alpha_{\rm
B}^{\prime}(r)$ is positive. In Fig.~4 the coupling $\alpha_{\rm
F}(r)$ together with $\alpha_{\rm B}(r)$ with parameters from
Eq.~(\ref{eq.18}) are plotted, which shows that $\alpha_{\rm F}(r)$
is smaller by $\sim 20\%$ in the region $0.5 - 0.6$~fm.

To compare our results with the existing lattice data we introduce
the dimensionless function $r^2 F_{\rm B}(r)$ and calculate two
characteristic scales: $r_1$ and $r_0$ \cite{ref.33}:
\begin{equation}
 r_1^2 F_{\rm B}(r_1)=1.0, \quad r_0^2 F_{\rm B}(r_0)=1.65,
\label{eq.27}
\end{equation}
where the function
\begin{equation}
 r^2 F_{\rm B}(r)= r^2 \sigma +\frac{4}{3}\alpha_{\rm F}(r),
\label{eq.28}
\end{equation}
depends on both $\sigma$ and $\alpha_{\rm F}(r)$.
For the static potential, like the Cornell and some lattice
potentials, with the coupling equal to a constant, one
has $\tilde V_{\rm st} (r) = \sigma r - \frac{e}{r}$ (where
$e=\frac{4}{3}\alpha_{\rm lat}=constant$) and therefore in the static
force,
\begin{equation}
 \tilde F(r) =\sigma +\frac{e}{r^2} ,
\label{eq.29}
\end{equation}
the coupling $\alpha_{\rm F}=\frac{3}{4}e$ is also constant.

\begin{figure}
 \includegraphics[width=60mm]{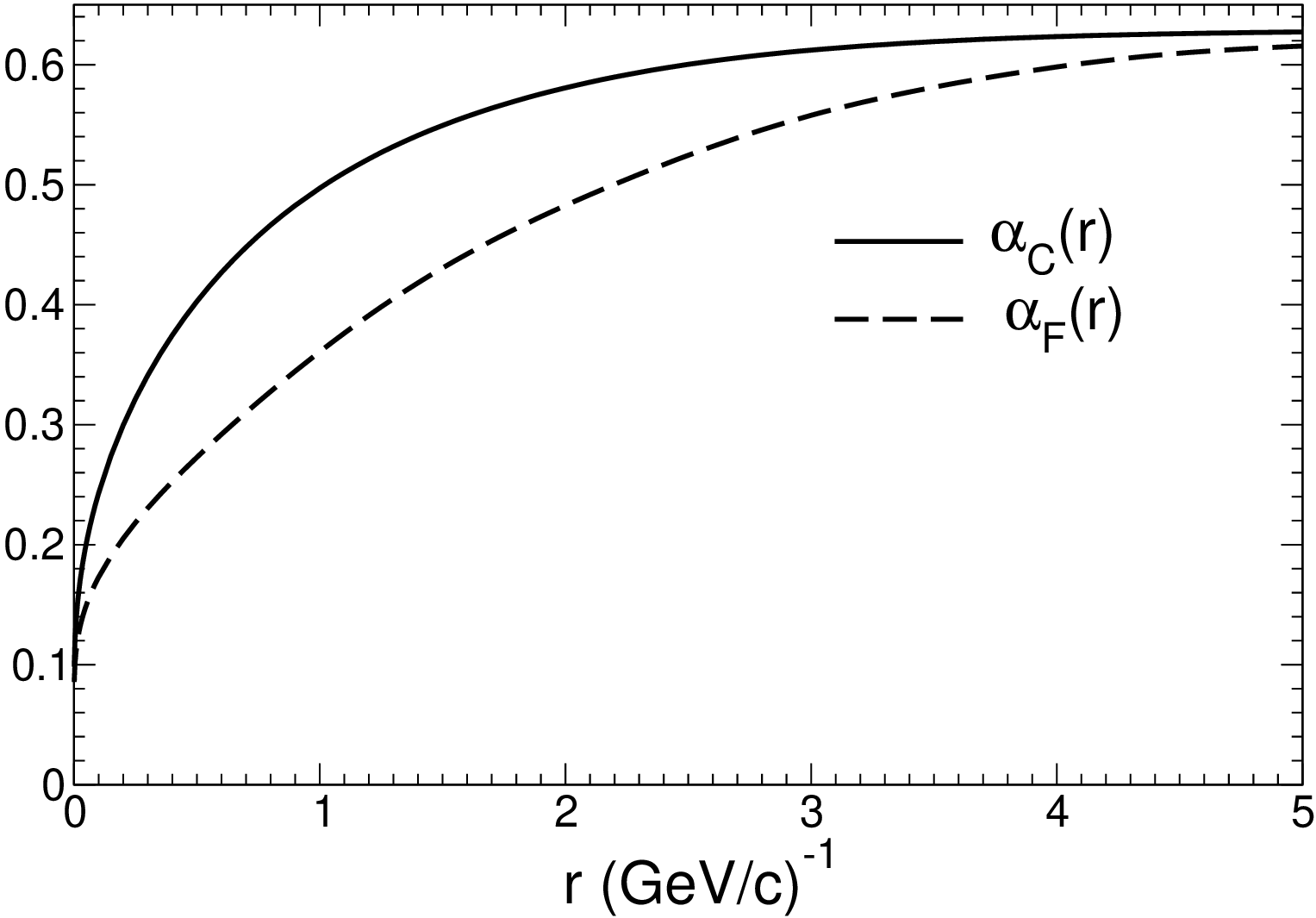}
\caption{The compound $\alpha_{\rm F}(r)$ and
 $\alpha_{\rm B}(r)$, taking  $\Lambda_{\rm V}(n_f)$ from Eq.~(\ref{eq.18})
 with $M_{\rm B}=1.15$~GeV, $\Lambda_{\rm V}(n_f=3)=0.4974$~GeV, and
 $\alpha_{{\rm B}~{\rm crit}}=0.63015$.
\label{fig.4}}
\end{figure}

On the contrary, in our calculations the coupling $\alpha_{\rm F}(r)$
changes rapidly in the region $0\leq r\leq 0.4$~fm, approaching
$\alpha_{\rm B}(r)$ only at large distances $r\geq 0.8$~fm (see
Fig.~\ref{fig.4}, where $\Lambda_V(n_f)$ is taken from Eq.~(\ref{eq.18})
with $M_{\rm B}=1.15$~GeV).

With the use of Eq.~(\ref{eq.27}) one easily calculates the
characteristic sizes:
\begin{equation}
  r_1=1.530~{\rm GeV}^{-1}=0.303~{\rm fm}, \quad
 r_0=2.321~{\rm GeV}^{-1}=0.460~{\rm fm},
\label{eq.30}
\end{equation}
taking $\Lambda_{\rm V}(n_f=3)=497.4$~MeV, which corresponds to
the central value of the perturbative $\Lambda_{\overline{\rm
MS}}(n_f=3)=337$~MeV. These numbers appear to be very close to
those calculated on the lattice: $r_1({\rm
lat})=1.621~$GeV$^{-1}=0.321$~fm \cite{ref.34} and
$r_0=1.3656(20)$~GeV$^{-1}=(0.468\pm 0.004)$~fm \cite{ref.27},
being only several percent smaller. However, to reach precise
agreement with the lattice scales one needs to take a bit smaller
$\Lambda_{\overline{\rm MS}}(n_f)$, namely, those values which
correspond to the lower bounds in pQCD (\ref{eq.4}):
$\Lambda_{\overline{\rm MS}}(n_f=5, 2-{\rm loop})=208$~MeV,
$\Lambda_{\overline{\rm MS}}(n_f=4)=279.2$~MeV,
$\Lambda_{\overline{\rm MS}}(n_f=3)=322$~MeV. For this choice we
have obtained $\alpha_{\rm crit}=0.5712$, $r_1=0.312$~fm, and
$r_0=0.470$~fm, which coincide with the lattice scales from
Refs.~\cite{ref.34, ref.35} with an accuracy better than $3\%$.

In the cases considered, we have found the following values for the product
$r_0~\Lambda_{\overline{\rm MS}}(n_f=3)$:
\begin{eqnarray}
 r_0~\Lambda_{\overline{\rm MS}}(n_f=3) & = & 0.764,
 ~~{\rm for}~~r_0=0.470~{\rm fm},~~\alpha_{\rm crit}=0.5712,
\nonumber \\
 r_0~\Lambda_{\overline{\rm MS}}(n_f=3) & = & 0.782,
 ~~{\rm for}~~r_0=0.460~{\rm fm},~~\alpha_{\rm crit}=0.630,
\label{eq.31}
\end{eqnarray}
in good agreement with the lattice results from Refs.
\cite{ref.27,ref.34}.
Thus we conclude that for large $\Lambda_{\rm V}(n_f=3)\sim 500$~MeV,
with the corresponding $\Lambda_{\overline{\rm MS}}=0.339$~MeV, one needs
to take a relatively large IR regulator $M_{\rm B}=1.15$~GeV to obtain the
scales $r_1,~r_0$ in agreement with the lattice results.
For a smaller regulator, e.g. $M_{\rm B}=1.0$~GeV, the scales
$r_1,~r_0$ turn out to be smaller: $r_1=0.292$~fm, $r_0=0.442$~fm,
giving $r_0~\Lambda_{\overline{\rm MS}}(n_f=3)=0.757$.
On the contrary, the large value $r_0=0.50$~fm may be obtained in
two ways: either with the larger regulator $M_{\rm B}\ga 1.30$~GeV, if
the ``perturbative" $\Lambda_{\overline{\rm MS}}(n_f=3)$ from
Eq.~(\ref{eq.4}) is used, or taking the significantly smaller
value of $\Lambda_{\overline{\rm MS}}(n_f=3)\sim 245$~MeV, as in
the quenched calculations \cite{ref.32}.
Thus we can conclude that the scale $r_0$ cannot be considered as
a universal parameter but depends on which values of $\Lambda$ and $M_{\rm B}$
are used. Notice, that the force $F_{\rm B}(r)$ depends also on the
string tension and our calculations with $\sigma=0.18$~GeV$^2$
give the scales $r_1$ and $r_0$ in good agreement with the lattice
results.

\begin{figure}[htb]
\begin{center}
 \includegraphics[width=60mm]{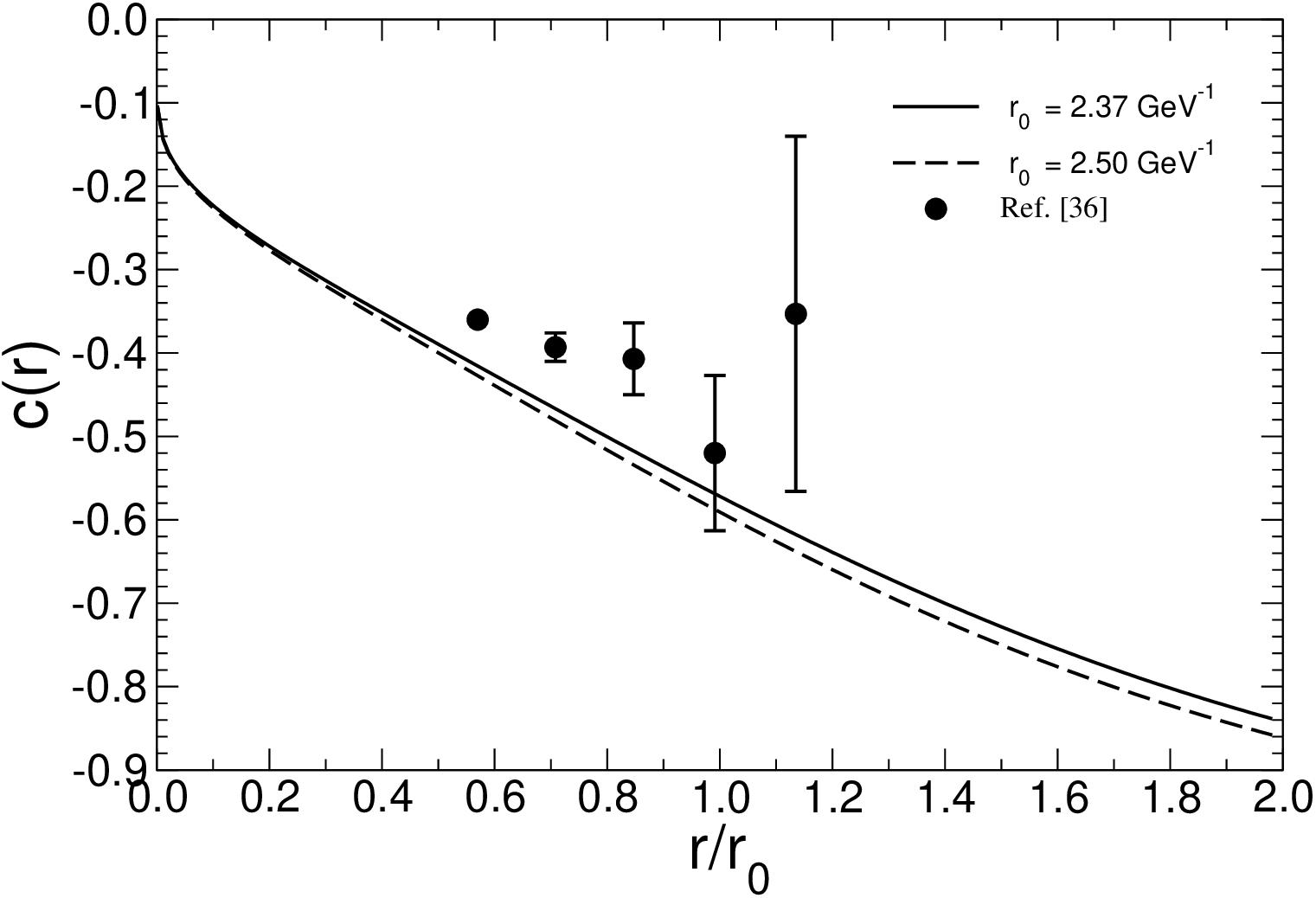}
\caption{The function $c(r)$ for $M_{\rm B}=1.15$ GeV and
$\Lambda_V(n_f)$ from Eq.~(\ref{eq.18}); the points with the
errors are taken from Ref.~\cite{ref.36}, where the function
$c(r)$ was calculated on the lattice with $n_f=2$. \label{fig.5}}
\end{center}
\end{figure}

An additional and very important test of the vector coupling comes
from the study of the function $c(r)$, which is defined via the
second derivative of the static potential and therefore does not
depend on the string tension. It is determined by $\alpha_{\rm
B}(r)$ and its first and second derivatives:
\begin{equation}
 c(r)= \frac{1}{2} r^3~V_{\rm st}^{\prime\prime}(r)=
 -\frac{4}{3}\alpha_{\rm F}(r)-
 \frac{4}{3} \alpha_{\rm {\rm B}}^{\prime\prime}(r) \frac{r^2}{2}.
\label{eq.32}
\end{equation}
The second derivative $\alpha_{\rm B}^{\prime\prime}(r)$ is
negative and therefore the magnitude of $c(r)$ appears to be
smaller than that of $\frac{4}{3} \alpha_{\rm F}(r)$. Moreover,
the slope of $c(r)$ depends on the IR regulator used.  The
behavior of $c(r)$ in our case, with $n_f=3$, is shown in
Fig.~\ref{fig.5} together with the points taken from the lattice
calculations of the ALPHA group, with $n_f=2$ \cite{ref.36}
(unfortunately, we could not find any lattice data in full QCD).
From Fig.~\ref{fig.5} one can see a qualitative agreement with both results.

\section{The bottomonium spectrum as a test of $\alpha_{\rm crit}$}
\label{sec.VII}

In bottomonium the centroid mass $\bar M(nl)$ for a given
multiplet $nl$ just coincides with the eigenvalue $\bar M(nl)$ of the
Hamiltonian given in Eq.~(\ref{eq.8}):
\begin{equation}
 \left[2\sqrt{\vep^2 + m_b^2} + V_{\rm st}(r)\right]\varphi_{nl}
 = \bar M(nl)\varphi_{nl} .
\label{eq.33}
\end{equation}
Our present calculations using the SSE (the relativistic
case) have better accuracy than in the nonrelativistic and
so-called einbein approximations, although the differences between
them are only $\sim 10$ MeV for low-lying masses $\bar M(nl)$ and
their mass splittings.

The bottomonium spectrum was calculated taking the compound
$\alpha_{\rm B}(r)$ with the values of $\Lambda_{\rm V}(n_f)$ from
(\ref{eq.18}) and also in the case with fixed $n_f=3$ and the same
$\Lambda_{\rm V}(n_f=3)=497.4$~MeV, but without matching. It
appears that in these two cases the differences between the masses
calculated are very small, $\sim 2-6$~ MeV. Therefore it is of
special importance to calculate the spectrum, considering
a different $\Lambda_{\rm V}$ for $n_f=3$, or a different
$\Lambda_{\overline{\rm MS}}(n_f=3)$. Our calculations show that
there are several mass splittings, which are most sensitive to the
choice of the ratio $\eta^2=\frac{M_{\rm B}^2}{\Lambda_{\rm
V}^2}$, determining the frozen value of $\alpha_{\rm B}(r)$. Their
experimental values are taken from Refs.~\cite{ref.18, ref.37,
ref.38}:
\begin{eqnarray}
 \bar{M}(2P) - \bar{M}(1P) & = & (360.0\pm 1.7)~{\rm MeV},
\nonumber \\
 \bar{M}(3P) - \bar{M}(2P) & = & (280\pm 14)~{\rm MeV},
\nonumber \\
 \bar{M}(1D) - \bar{M}(1P) & = & (264\pm 2)~{\rm MeV}.
\label{eq.34}
\end{eqnarray}
Here the centroid mass  $\bar M(\chi_b(3P))\simeq (10540\pm
5)$~MeV is estimated from two experimental masses measured by the
ATLAS \cite{ref.37} and the D0 \cite{ref.38} collaborations.

In Ref.~\cite{ref.13} it was already demonstrated that the fit to
the bottomonium splittings appears to be sensitive to the choice
of the critical coupling constant. However, in the vector coupling
the parameters were often taken in a rather arbitrary way. In
particular, for the lattice static potential with small
$\alpha_{\rm lat}(r)=const=0.306$ ($n_f=3$) the $1D-1P$ and
$1P-1S$ splittings are by $40-30$~MeV smaller than their
experimental values.

Here, we first determine the masses of the $1D$ multiplet. The
fine-structure splittings of this multiplet were calculated taking
the strong coupling in the spin-orbit and tensor interactions close
to the value for the $1P$-states \cite{ref.30}, namely,
$\alpha_{\rm FS}(\mu_{\rm FS})=0.40$ at the scale $\mu_{\rm
FS}\sim 1$~GeV. We take the spin-orbit splitting $a(1D)$ and
tensor splitting $t(1D)$ in one-loop approximation \cite{ref.39}:
\begin{eqnarray}
 a(nl) = \frac{2\alpha_{\rm FS}}{\omega(nl)^2}\langle r^{-3}
 \rangle_{nl} - \frac{\sigma}{2\omega(nl)^2}\langle r^{-1}\rangle_{nl},\quad
 c(nl)=\frac{4\alpha_{\rm FS}}{3\omega(nl)^2}\langle r^{-3}\rangle_{nl},
\label{eq.35}
\end{eqnarray}
where in the denominator, instead of the quark mass usually used
\cite{ref.40}, one has to put the quark kinetic energy. This
result follows from the analysis of the spin-dependent part of the
RSH $H_{\rm R}$ in Ref.~\cite{ref.39}. For the $1D$ states the
calculated kinetic energy is $\omega(1D)=5.0$~GeV.
For the $1D$ states and the set of the parameters from Table 1
with $M_{\rm B}=1.15$~GeV, the following values of the matrix
elements were calculated: $\langle r^{-3}\rangle_{\rm
1D}=0.132(2)$~GeV$^3$ and $\langle r^{-1}\rangle_{\rm
1D}=0.444(4)$~GeV. Then the fine-structure splittings
$a(1D)=2.6$~MeV and $c(1D)=2.8$~MeV were calculated. Then the
masses of the $n\,^3D_J$ states are defined as in
Ref.~\cite{ref.40},
\begin{eqnarray}
 M(1\,^3D_3) & = & \bar M(1D) +2 a -\frac{c}{7},
 \nonumber \\
 M(1\,^3D_2) & = & \bar M(1D) -~a + \frac{c}{2},
 \nonumber \\
 M(1\, ^3D_1) & =  & \bar M(1D)-3 a -\frac{c}{2}.
\label{eq.36}
\end{eqnarray}
Then taking from the experiments the centroid mass $\bar
M(1D)=10164$~MeV (see below) and the values of $a$ and $c$, using
Eq.~(\ref{eq.35}), the following masses $M(n\,^3D_J)$ are
obtained:
\begin{eqnarray}
 M(1\,^3D_3)=(10169\pm 2)~{\rm MeV},
\nonumber \\
 M(1\,^3D_2)=(10163\pm 2)~{\rm MeV},
\nonumber \\
 M(1\,3D_1)=(10155\pm 2)~{\rm MeV}.
\label{eq.37}
\end{eqnarray}
The centroid mass used, according to (\ref{eq.36}), is by $\sim
1$~MeV larger than $M(1\,^3D_2)=(10163\pm 2)$~MeV, known from
experiments, i.e., $\bar M(1D)=(10164\pm 2)$~MeV. From
Eq.~(\ref{eq.37}) our calculations give the splittings of the $1D$
multiplet: $M(1\,^3D_3)-M(1\,^3D_1)=14(4)$~MeV and $M(1\,^3D_2)-
M(1\,^3D_1)=8(4)$~MeV, which appear to be two times smaller than
those obtained in lattice calculations \cite{ref.41}.

For the $nP$ bottomonium multiplets their spin-averaged masses are
known very accurately \cite{ref.18,ref.42,ref.43}:
\begin{equation}
 \bar{M}(1P)=(9900.0\pm 0.6)~{\rm MeV}, \quad
 \bar{M}(2P)=(10260\pm 0.7)~{\rm MeV},
\label{eq.38}
\end{equation}
and therefore their mass splitting $\bar M(2P) - \bar
M(1P)=260(2)$~MeV is also known with great accuracy and may be
used as a test of different sets of the parameters.

\begin{table}
\caption{\label{tab.1} The mass splittings in bottomonium in MeV
($\Lambda_{\overline{\rm MS}}(n=3)=325$~MeV)}
\begin{tabular}{|c|c|c|c|}
\hline
       & $M_{\rm B}=1.15$~GeV  & $M_{\rm B}=1.10$~GeV  & ~exp.~\cite{ref.18}~ \\

   State   & $m_b=4.832$~GeV & $m_b=4.840$~GeV &                   \\
\hline
 ~$1D-1P$~ &  259            &  261            &        $264\pm 2$ \\
 ~$2P-1P$~ &  371            &  376            &        $360\pm 2$\\
 ~$3P-2P$~ &  288            &  294            &        $280\pm 14$\\
\hline
\end{tabular}
\end{table}

As seen from Table~\ref{tab.1}, the $1D-1P$ splitting is in good
agreement with experiment for $M_{\rm B}=1.10$~GeV, however, at
the same time the $2P-1P$ splitting increases with decreasing
$M_{\rm B}$ and one needs to reach the best agreement for both
splittings. Choosing different sets of the parameters $M_{\rm B}$
and $\Lambda_V(n=3)$, we have observed that a good agreement with
the experimental splittings takes place for the frozen constant
$\alpha_{\rm crit}=0.57\pm 0.02$ and the IR regulator $M_{\rm
B}=(1.15\pm 0.02)$~GeV, while the calculated spin-averaged masses
coincide with experiment within $\pm (5-10)$~MeV. Notice, that the
lattice calculations give larger $D$-wave masses \cite{ref.41}, as
compared to ours, while smaller masses $M(1D)$ were predicted in
Ref.~\cite{ref.40}.

For the $2P-1P$ splitting a small deviation $\sim 5$~MeV from the
experimental value can be obtained, if a smaller QCD constant
$\Lambda_{\overline{\rm MS}}(n=3)=(317\pm 5)$~MeV is used, while
$M_{\rm B}=(1.15\pm 0.02)$~GeV is relatively large.

We do not give here the centroid masses of the $S$-states, because
for calculations of $\bar M(1S),~\bar M(2S)$ one needs to take
into account the nonlinear behavior of the confining potential at
short distances, $r\lsim 0.20$~fm, and this fact gives rise to an
additional uncertainty -- small negative corrections to the
$S$-wave masses. Such corrections are very small for the states
with $l\neq 0$, since their w.f.s are equal to zero near the
origin, while the $S$-wave w.f.s have their maximum values there.

\section{Conclusions}
\label{sec.VIII}
We have studied the vector coupling in the momentum and coordinate
spaces, introducing the IR regulator, $M_{\rm
B}=\sqrt{2\pi\sigma}=(1.06\pm 0.11)$~GeV, as it is prescribed in
BPT.

For the vector coupling in momentum space we have performed the
matching procedure at the quark mass thresholds and calculated the
``vector" constants $\Lambda_{\rm V}(n_f)$. It appears that these
constants correspond to $\Lambda_{\overline{\rm MS}}(n_f)$, which
coincide with the perturbative $\Lambda_{\overline{\rm MS}}(n_f)$
within $\pm 5$~MeV. Moreover, their values weakly depend on the
regulator $M_{\rm B}$, if it is taken from the range
$(1.0-1.20)$~GeV.
We have shown that in the static force the scales $r_0$ and $r_1$
are not universal numbers and depend on the IR regulator used. Thus
$r_0$ decreases by $6\%$ when $M_{\rm B}$ decreases from the value
$M_{\rm B}=1.15$~GeV to $M_{\rm B}=1.00$~GeV.

The ratio $\frac{r_0}{r_1}=1.505\pm 0.02$ and the product $r_0
\Lambda_{\overline{\rm MS}}(n_f=3)=0.77\pm 0.02$ were calculated.
The choice with $M_{\rm B}=1.15$~GeV and $\Lambda_{\overline{\rm
MS}}(n_f=3)=322$~MeV gives the best (precise) agreement with the
lattice scales $r_0$ and $r_1$.
The function $c(r)$, which is  proportional to the second
derivative of the static potential and does not depend on the
string tension, is calculated. This function illustrates that the
saturation of the vector coupling takes place at the distances
$r\ga 0.6$~fm.

Our analysis of the bottomonium spectrum shows that the splittings
$\bar M(1D)-\bar M(1P)$ and $\bar M(2P)-\bar M(1P)$ are very
sensitive to the factor  $\eta^2=\frac{M_{\rm B}^2}{\Lambda_{\rm
V}(n_f=3)^2}$ and the best agreement with experiment is reached
taking $\eta=2.46\pm 0.04$.
We have derived the value of the  frozen coupling: $\alpha_{\rm
crit}=0.57\pm 0.02$.
The following splittings for the members of the bottomonium $1D$
multiplet are predicted:
$M(1\,^3D_3)-M(1\,^3D_1)=14(2)$~MeV,~~$M(1\,^3D_2)-M(1\,^3D_1)=8(2)$~MeV.

\acknowledgments

The authors are grateful to Yu.~A.~Simonov for useful discussions
and suggestions.


\begin{thebibliography}{99}

\bibitem{ref.1}
E.~Eichten, K.~Gotfried, T.~Kinoshita, K.~D. Lane, T.M.~Yan, 
Phys. Rev. D {\bf 21}, 203 (1980); Phys. Rev. Lett. {\bf 34}, 369 (1975).

\bibitem{ref.2}
D.~P.~Stanley and D.~Robson, Phys. Rev. D {\bf 21}, 3180 (1980);
Phys. Rev. Lett. {\bf 45}, 235 (1980); L.~P.~Fulcher, Phys. Rev. D
{\bf 44}, 2079 (1991); ibid D {\bf 50}, 447 (1994);
J.~L.~Basdevant and S.~Boukraa, Z. Phys. C {\bf 28} (1983).

\bibitem{ref.3}
S.~Godfrey and N.~Isgur, Phys. Rev. D {\bf 32}, 189 (1985).

\bibitem{ref.4}
D.~Ebert, R.~N.~Faustov, and V.~O.~Galkin, Phys. Rev. D {\bf 67},
014027 (2003); Mod. Phys. Lett. A {\bf 20}, 1887 (2005) .

\bibitem{ref.5}
W.~Lucha, F.~F.~Schoberl, and D.~Gromes, Phys. Rept. {\bf 200},
127 (1991) and references therein.

\bibitem{ref.6} A.~Yu.~Dubin, A.~B.~Kaidalov, and
Yu.~A.~Simonov, Phys. Lett. B {\bf 323}, 41 (1994); Yad. Fiz. {\bf
56}, 213 (1993); E.~Gubankova and A.~Yu.~Dubin, Phys. Lett. B {\bf
334}, 180 (1994) .

\bibitem{ref.7}
Yu.~A.~Simonov, Nucl. Phys. B {\bf 307}, 512 (1988); G.~Dosch and
Yu.~A.~Simonov, Phys. Lett. B {\bf 205}, 339 (1988);
Z. Phys. C {\bf 45}, 147 (1989).

\bibitem{ref.8}
A.~DiGiacomo, H.~G.~Dosch, V.~I.~Shevchenko, and Yu.~A.~Simonov,
Phys. Rept. {\bf 372}, 319 (2002).

\bibitem{ref.9}
A.~M.~Badalian and  B.~L.~G.~Bakker, Phys. Rev. D {\bf 66}, 034025
(2002).

\bibitem{ref.10}
A.~M.~Badalian and V.~P.~Yurov, Phys. Atom. Nucl. {\bf 56}, 176
(1993); Sov. J. Nucl. Phys. {\bf 51}, 869 (1990)

\bibitem{ref.11}
V.~Shevchenko and Yu.~A.~Simonov, Phys. Rev. Lett. {\bf 85}, 1811
(2000); hep-ph/0104135 (2001).

\bibitem{ref.12}
W.~W.~Repko, S.~F.~Radford, and M.~D.~Santio, arXiv:1211.6373
[hep-ph] and references therein.

\bibitem{ref.13}
A.~M.~Badalian, A.~I.~Veselov, and B.~L.~G.~Bakker, Phys. Rev. D
{\bf 70}, 016007 (2004).

\bibitem{ref.14}
C.~W.~Bernard et al. (MILK Collab.), Phys. Rev. D {\bf 64}, 054506
(2001); ibid. Phys. Rev. D {\bf 62}, 034503 (2000).

\bibitem{ref.15}
Y.~Koma and M.~Koma, arXiv:1211.6795; Nucl. Phys. B {\bf 769}, 79
(2007).

\bibitem{ref.16}
Yu.~A.~Simonov,  Phys. Atom. Nucl. {\bf 74}, 1223 (2011);
arXiv:1011.5386 (2010)[hep-ph].

\bibitem{ref.17}
M.~Peter, Phys. Rev. Lett. {\bf 78}, 602 (1997); Nucl. Phys. B
{\bf 501}, 471 (1997); M.~Jezabek, M.~Peter, and Y.~Sumino, Phys.
Lett. B {\bf 428}, 352 (1998); Y.~Schr\"{o}der Phys. Lett. B {\bf
447}, 321 (1999).

\bibitem{ref.18}
J.~Beringer et al. (Particle Data Group), Phys. Rev. D {\bf 86},
010001 (2012).

\bibitem{ref.19}
S.~Bethke, arXiv:1210.0325 (2012) [hep-ex]; Eur. Phys. J. C {\bf
64}, 689 (2009); arXiv:1210.0324 [hep-ex] and references therein.

\bibitem{ref.20}
A.~M.~Badalian and D.~S.~Kuzmenko, Phys. Rev. D {\bf 65}, 016004
(2001); A.~M.~Badalian, Phys. Atom. Nucl. {\bf 63}, 2173 (2000).
(2002)

\bibitem{ref.21}
G.~Bali, Phys. Lett. B {\bf 460}, 170 (1999).

\bibitem{ref.22}
D.~V.~Shirkov, arXiv:1208.2103 (2012) [hep-ph].

\bibitem{ref.23}
G.~S.~Bali, Phys. Rev. D {\bf 62}, 114503  (2000).

\bibitem{ref.24}
E.~J.~Eichten and C.~Quigg, Phys. Rev. d {\bf 52}, 1726 (1995).

\bibitem{ref.25}
A.~M.~Badalian, B.~L.~G.~Bakker, I.V. Danilkin, Phys. Rev. D
{\bf 79}, 037505 (2009); Phys. Atom. Nucl. {\bf 73}, 138 (2010).

\bibitem{ref.26}
G.~Parisi and R.~Petronzio, Phys. Lett. B {\bf 94}, 51 (1980);
J.~M.~Cornwall, Phys. Rev. D {\bf 26}, 1453 (1982);
A.~C.~Mattingly and P.~M.~Stevenson, Phys. Rev. D {\bf 49}, 437
(1994).

\bibitem{ref.27}
A.~Bazavov et al., arXiv:1205.6155 [hep-lat]; Phys. Rev. D {\bf
85}, 054503 (2012) and references therein; N.~Brambilla,
X.~Garcia i Tormo, J.~Soto, and A.~Vairo, Phys. Rev. Lett. {\bf 105},
212001 (2010).

\bibitem{ref.28}
A.~M.~Badalian, B.~L.~G.~Bakker, and I.~V.~Danilkin, Phys. Rev. D
{\bf 81}, 071502 (2010); Erratum-ibid: D {\bf 81}, 099902 (2010);
Phys. Atom. Nucl. {\bf 74}, 631 (2011).

\bibitem{ref.29}
Yu.~A.~Simonov, Phys. Lett. B {\bf 515}, 137 (2001).

\bibitem{ref.30}
A.~M.~Badalian and  B.~L.~G.~Bakker,  Phys. Rev. D {\bf 62},
094031 (2000).

\bibitem{ref.31}
C.~T.~H.~Davies, E. Follana, I.D. Kendall, G.P. Lepage, C. McNeile,
Phys. Rev. D {\bf 81}, 034506 (2010);
M.~Cheng et al., Phys. Rev. D {\bf 81}, 054504 (2010).

\bibitem{ref.32} S.~Capitani, M.~Luescher, R.~Sommer, and
H.~Wittig, Nucl. Phys. B {\bf 544}, 669 (1999).

\bibitem{ref.33}
R.~Sommer, Nucl. Phys. B {\bf 411}, 839 (1994).

\bibitem{ref.34}
A.~Gray et al., Phys. Rev. D {\bf 72}, 094507 (2005).

\bibitem{ref.35}
R.~J.~Dowdall et al. (HPQCD Collab.), arXiv:1110.6887 (2011)
[hep-lat].

\bibitem{ref.36}
M.~Donnellan  et al., Nucl. Phys. B {\bf 849}, 45 (2011) and
references therein.

\bibitem{ref.37}
G.~Aad et al. (ATLAS Collab.), Phys. Rev. Lett. {\bf 108}, 152001
(2012).

\bibitem{ref.38}
V.~M.~Abazov et al. (D0 Collab), Phys. Rev. D {\bf 86}, 031103
(2012).

\bibitem{ref.39}
A.~M.~Badalian, A.~V.~Nefediev, and Yu.~A.~Simonov, Phys. Rev. D
{\bf 78}, 114020 (2008).

\bibitem{ref.40}
W.~Kwong and J.~L.~Rosner, Phys. Rev. D {\bf 38}, 279 (1988).

\bibitem{ref.41}
J.~O.~Daldrop, C.~T.~H.~Davies, and R.~J.~Dowdall,
arXiv:1112.2590 (2011) [hep-lat].

\bibitem{ref.42}
G.~Bonvicini et al. (CLEO Collab.), Phys. Rev. D {\bf 70}, 032001
(2004).

\bibitem{ref.43}
P.~Del Amo Sanchez et al. (BaBar Collab.), Phys. Rev. D {\bf
82}, 111102 (R) (2010).

\end{thebibliography}
\end{document}